\documentstyle[aps,12pt]{revtex}

\begin{document}
\begin{title}
{Coulomb gas, dipoles and a generalization of the Debye-Hukkel approximation 
through the path integral representation}
\end{title}
 \author{D.Yu.Irz}
\address{Institute for
High Pressure Physics, Russian Academy of Sciences, Troitsk 142092, Moscow
region, Russian Federation}
\date{\today}
\maketitle

\begin{abstract}
The Coulomb gas partition sum is rewritten in terms of the path integrals 
formalism. It is shown that perturbation theories based on the Mayer 
expansion and on the path integrals method lead to the identical results.
The well known Debye-Hukkel result for the case of 3D Coulomb plasma
is completely rederived. An analogous result is obtained for the case of
the Coulomb gas with dipoles. This result can be considered as a
generalization of the Debye-Hukkel approximation. Other possible
generalizations including the range of interaction potentials for which
the Debye-Hukkel approximation can be applyed are discussed.
\end{abstract}
\section{Introduction}
\label{intro}

The problem of calculating of the Gibbs partition sum is well known
since the pioneering works in classical satatistical mechanics.
The simplest example of precise calculaton of the Gibbs free energy
is the classical ideal gas. The more general problem of classical
statistical
mechanics is the evaluation of partition sum for  a nonideal system.
The main difficulty is the calculation of the configuration integral
$$
Q=\int dr_1\dots dr_N e^{-\beta U(r_1\dots r_N) }.
$$

This problem has no explicit solution.   There are several approaches
to analyse the thermodynamics of non-ideal systems with short range
interactions:
Mayer's method, correlation function method, integral equation method,
renormalization group method  e.t.c.~\cite{balesku}

Some  additional difficulties appear while analysing Coulomb-like systems
because of the long range type of the interection.
Since the pioneering work of Debye,
the standard solution of this problem is the Debye-Hukkel (DH) mean field
approximation~\cite{balesku,mayer} where the  Debye-Hukkel screened
interaction appears instead of Coulomb one.
This method  is widely used in different fields of physics to analyze 3D
plasma-like systems.

Howewer, the extention of DH theory to the case of more complex
systems is still a challenge to the statistical mechanics.
On our opinion, the main reason for
this is the absence of the consistent point of view on DH approximation.
Really, a number of ways to obtain  DH result are known.
First of all this is the linearization of the mean field equations which is
the simplest and well known way to obtain DH screening.
Another way to obtain the DH result is a cycle
approximation~\cite{balesku} applied to the Mayer's diagrams. Howewer, like
other chain approximations in statistical mechanics, this method leaves open
the question how to choose the relevant chain class.
We will discuss  another method based on the special property of the
configuration integral and the analogy with multiple Gaussian
integrals.~\cite{samuel}

Despite the fact that the first attempt to apply method discussed below
to continuous systems was undertaken in 1959 \cite{edwards}, to our knowlege
this approach has never been  used as a regular method to analyse classical
systems. At the same time some exotic relations between different correlation
functions were
obtained by using this approach (See for example~\cite{minnh87}).
The analogous method
has been used to prove the equivalence of a 2D Coulomb gas model to the
Euclidean Sine-Gordon model~\cite{samuel}.
This equivalence was considered as a tool allowing to obtain some new results
in quantum field theory from thermodynamics properties of the 2D Coulomb gas.

There exists some papers in statistical mechanics~\cite{kholodenko},
where authors try to obtain some new results for  Coulomb systems
by using the
Sine-Gordon analogy and the technique
of the quantum field theory. But due to some difficulties some
well known results for the 3D plasma were not completely rederived
in these papers.

Some purely mathematical papers published in 80-th should be specially
mentioned~\cite{math}. In these papers the 3D Coulomb
system has been analysed on the base of its equivalence to the 3D
Sine-Gordon model. A strict mathematical theory was developed for
this case.
As it was shown, the Debye-Hukkel (DH) result can be easily
reproduced in the  lowest order of this approach.
This is the simplest and the most consistent method to obtain the well
known DH result.

In those papers, however, the emphasis was made on the mathematical
problems, and it is difficult to see the physical sense of the
tranformations.
Also there are no discussions on possible extension of this
method to the case of more complex non-Coulombic systems.
The present paper is an attempt to overcome some of these difficulties and
to present the application of  Gaussian integral method
to continuous systems.

In the section \ref{gau} the basis of the Gaussian integral formalism
applied to the models of classical statistical mechanics is described.
Section \ref{diagr} is dedicated to the perturbation theory, that is fully
analogous to the usual quantum field theory's one. The correspondence
of this theory to the usual Mayer's expansion in powers of the
activity $z$ is analysed in detail.
Some mathematical restrictionscaused by the features of our method
are also considered. As it is shown, these restrictions are strong enough.
This explains, why the DH-like mean fied theory can not be applied in the
case of non-Coulomb systems.

 In the section \ref{density} the expansion in  powers of density
is presented. It is shown that the lowest order of this expansion
gives precisely the DH approximation. The exact reproduction
of this well known result permits us to  expect that the
method used in this paper can lead us to some new results in the case of
more complexes systems. One such example is given in
section~\ref{dipols}. The results of the present paper are summarized
in~\ref{conclusion}. Some advantages and  defects of the method
are reviewed  in the conclusion.

\section{Gaussian Formalism for the continuum case}
\label{gau}

Let us consider the functional:
\begin{equation}
\label{gauss}\frac 1{{\cal N}}\int \,D\chi (r)\exp \left\{ -\frac \beta
2\int \chi (r)drH(r-r')dr'\chi (r')+i\beta \int
\rho (r)\chi (r)dr\right\} =I[\rho ],
\end{equation}
where:
$$
{\cal N}=\int \,D\chi (r)\exp \left\{ -\frac \beta 2\int \,\chi
(r)drH(r-r')dr'\chi (r')\right\}
$$
-- is a normalizing factor. $H(r)$ is a kernel of some linear operator. $%
D\chi (r)$ is a mesure on the space of real functions. In other words, in
Eq.(\ref{gauss}) we integrate over all pathes $\chi (r)$ as one usually do in
quantum field theory \cite{ramon} and quantum mechanics.

Now we can introduce a function $V(r)$ satisfying  the condition:
\begin{equation}
\label{Hdef}\int \,V(r-r')dr'H(r'-r'')=\delta (r-r'').
\end{equation}
This condition means that $V$ is the inverse operator relative to $H$.
(Here and later in this paper we suppose $H$ and $V$ to be translationally
invariant)

With the help of $V(r-r')$ we can transform (\ref{gauss}) to the form:
$$
\exp \left\{ -\beta /2\int \rho (r)drV(r-r')dr'\rho
(r')\right\} =I[\rho (r)]=\qquad \qquad \qquad \nonumber
$$
\begin{equation}
\label{main}
\qquad \int \frac{D\chi }{{\cal N}}\exp \left\{ \beta /2\int
\chi (r)drH(r-r')dr'\chi (r')+i\beta \int \,\rho
(r)\chi (r)dr\right\}.
\end{equation}
Later we shall examine the conditions under which the above transformation is
valid.

The main aim of this section is to associate  Eq.(\ref{main}) with the statistical
mechanics of a classical system. For this purpose we consider a system
consisting of $N=N_{+}+N_{-}$ charged point-like particles.
The microscopic charge density of the  system can be written as:
\begin{equation}
\label{rho}
\rho _q(r)=\sum_{i=1}^{N_{+}}\delta
(r-r_i)+\sum_{j=1}^{N_{-}}(-1)\delta (r-r_j).
\end{equation}
We also suppose that the interparticle interaction is central --
$s_is_jV(r_{ij})$ (where $s_i$, $s_j$ -- are the charges of particles),
so that we get for the full energy:
$$
U(r_1\dots r_N)=\frac 12\sum_{i\ne j=1}^{N_{+}}V(r_i-r_j)+\frac 12\sum_{i\ne
j=1}^{N_{-}}V(r_i-r_j)-\sum_{i=1}^{N_{+}}\sum_{j=1}^{N_-}V(r_i-r_j)=
$$
\begin{equation}
\label{gibbsen}
\int \rho _q(r)V(r-r')\rho _q(r^{\prime
})drdr'-(N_{+}+N_{-})V(0)
\end{equation}

Using Eq.~(\ref{Hdef}) and Eq.~(\ref{main}), we can write Gibbs factor
$\exp(-\beta U(r_1\dots r_N))$ of the system through the path integrals:
\begin{equation}
\label{gibbsgauss}
e^{-\frac{\beta}{2}\left( \sum\limits_{i\ne
j=1}^{N_+}V(r_i-r_j)+ \sum\limits_{i\ne j=1}^{N_-}V(r_i-r_j)-
2\sum\limits_{i=1}^{N_+}\sum\limits_{j=1}^{N_-}V(r_i-r_j)\right)}= e^{N\beta
V(0)/2}\int \frac{D\chi}{{\cal N}} e^{-\beta/2(\chi,H,\chi)+i\beta(\rho_q%
\chi)}.
\end{equation}
 Here we denoted for simplicity:
$$
(\chi,H,\chi)=\int\,\chi(r)drH(r-r^{\prime})dr^{\prime}\chi(r^{\prime}),\\
$$
$$
(\chi,\rho_q)=\int\,\chi(r)\rho_q(r)dr
$$
 Equation~(\ref{gibbsgauss}) will be the main for our later examinations
in this paper.
Applying analytical techniques  of path integrals calculations to
the Eq.~(\ref{gibbsgauss}),
we hope to receive {\it mathematically based} results for
thermodynamics of the system in question.
Talking about {\it mathematically based results } we mean the results
received directly from the basic physical formulas, using usual
mathematical ideas {\it without} additional physical supposition.
Our purpose is to demonstrate the main way to obtain physical
results, while the details of a mathematical consideration is the topic of interest
for mathematicians.

       Let us note the following:

{\bf a})  Interparticle potentials $V(r)$ ordinary contain
sigularities at $r\to 0$. Moreover, there is a large and very important
class of potentials weakly decreasing at $r\to\infty$. (Coulomb
potential in $d=2,3$ is the simplest example)  Due to these effects
divergences can appear in $H(r)$ and its  integrals.
(See Eq. (\ref{gibbsgauss})) This may
cause each component in the right hand side
of Eq.(\ref{gibbsgauss}) to be infinite. On this stage of consideration
we shall ignore this fact by using for our purposes the most general
notation.  For mathematical completeness we can suppose all divergences
caused by the form of $V(r)$ to be somehow regularized.
The regularization parameters should be taken equal to $0$ at the
final stage  of calculations after taking the usual limit
$N\to\infty$,
$\Omega\to\infty$, $\rho=const$.  We suppose that the final result
does not depend on the type of regularization.

{\bf b}) Let us consider the range of the applicability of
Eq. (\ref{gibbsgauss}).
Despite a seeming generality, this formula has  some very important
restrictions. To understand them we can write the discrete
matrix analogue of Eq. (\ref{main}):
\begin{equation}
\label{discr}
e^{\frac{1}{2}\sum\limits_{ij}\xi_iA^{-1}_{ij}\xi_j}= \frac{%
\displaystyle \int\,dx_1\dots dx_Ne^{ -\frac{1}{2}\sum%
\limits_{ij}x_iA_{ij}x_j +i\sum\limits_k\xi_kx_k } } {\displaystyle
\int\,dx_1\dots dx_Ne^{ -\frac{1}{2}\sum\limits_{ij}x_iA_{ij}x_j} }.
\end{equation}
It is wel known that this equation is correct only if  $||A_{ij}||$ is a
positively definite form. Otherwise integrals in the numerator and
the denominator of Eq. (\ref{discr}) are diverging.
Therefore Eq. (\ref{gibbsgauss}) is correct only for positively
definite operators $H(r-r')$. Due to the translational invariance of $H$,
we can write this condition with the help of Fourier transforms as:
\begin{eqnarray}
\hat H(q)&=&\int\,d{\bf r}e^{i{\bf qr}}H(r)>0;\nonumber\\
\hat V(q)&=&\frac{1}{\hat H(q)}>0.
\label{fourierV}
\end{eqnarray}
 We suppose here that the interparticle interaction
$V(r-r')=V(|{\bf r}-{\bf r}'|)$ is central, so that
$\hat V(q)=\hat V (-q)=\hat V^*(q)$

Coulomb potential is the simplest example satisfying (\ref{fourierV}):
\begin{equation}
\label{coulomb}
\hat{V}(q)=\frac{e^2}{q^2},\qquad d=2,3;
\end{equation}
Yukava (screened Coulomb) potential is the second example:
\begin{equation}
\hat{V}(q)=\frac{e^2}{q^2+\lambda^2},\qquad d=2,3;
\end{equation}

Another (slightly exotic) example is the intervortex interaction in a thin
superconducting film \cite{pearl}:
\begin{equation}
\hat{V}(q)=\frac{\varphi_0^2}{2\pi\Lambda}\frac{1}{q(q+1/\Lambda)},\qquad d=2
\end{equation}
 Here $\varphi_0=hc/2e$ is the flux quantum, $\Lambda$ is a
penetration depth
in two dimensions. Statistical mechanics of vortices
interacting via this potential is analysed in
\cite{pr2,pr1}.

{\bf c}) One should pay the attention to the factor $e^{\frac{N\beta}{2}V(0)}$
in Eq. (\ref{gibbsgauss}).
As far as we know, the most authors  \cite{samuel,minnh87,kholodenko} ignore
this factor.
To exclude it the potential
\begin{equation}
\label{edwrds}
\tilde V(r)=\left\{{1/r,\qquad r>\sigma}\atop{0,\qquad r<\sigma}\right.
\end{equation}
was used by Edwards~\cite{edwards} instead of Coulomb potential ($d=3$).
It is easy to show that
Fourier transform of Eq. (\ref{edwrds}) is not positively definite,
so that the correctness of (\ref{gibbsgauss}) is doubtful.

The most strict method is proposed in mathematical  papers (see for
example \cite{math}), where the interaction:
\begin{equation}
\label{rigour}
V_R(r)=\frac{1}{r}-\varepsilon\frac{e^{-r/\varepsilon}}{r}
\end{equation}
is used, instead of Coulomb potential.
The final result should be received after taking the limit $\varepsilon\to 0$.
This provides $V(q)$ to be positive, and $V(r)$ equal to zero as $r\to 0$,
so that our factor $\sim e^{V_r(0)}$ is equal to 1.

It is easy to see that Eq. (\ref{rigour}) is no more than one of different
ways to regularize $V(r)$. We can consider a more general method to smooth
singularities in $V(r)$:
$$
V_R(r_1-r_2)=\int\,dr^{\prime}dr^{\prime\prime}n(r_1-r^{\prime})V(r^{%
\prime}-r^{\prime\prime})n(r^{\prime\prime}-r_2),
$$
with $V_R(0)\ne 0$, so that the factor $e^{N\beta V(0)/2}$ cannot
be omited. We shall see later that this coefficient is very important in
the mean field and perturbation theories, based on Eq.(\ref{gibbsgauss}).

\section{Diagrammatic expansions}
\label{diagr}

To apply the Gauss-analogy (\ref{main}) to the problems of statistical
mechanics let us
consider a system consisting of "positively and negatively charged"
particles, having absolute activities $z_+$ and $z_-$.
The term "charged particles" is employed here in unusual sense:
The interparticle interaction may be not Coulomb, but it depends on
some special charge-like characteristics of particles. For example
the interaction between two vortices in a supercondoctor depends on
mutual direction of their magnetic fields.

The grand partition function for such a "plasma" at a temperature
$1/\beta$ can be written as:
\begin{equation}
\label{GrandPartSum} \zeta=\sum_{N_+,N_-=0}^{\infty}\frac{z_+^{N_+}z_-^{N_-}
}{N_+!N_-!} \int\{dr^{(+)}\}_{N_+}\{dr^{(-)}\}_{N_+} \exp\left\{ \frac{\beta
}{2}\sum_{i\ne j}s_is_jV(r_{ij}) \right\}.
\end{equation}
Using Eq.(\ref{main}) (or Eq. (\ref{gibbsgauss})) we can rewrite the
above equation:
\begin{equation}
\zeta=\sum_{N_+,N_-=0}^{\infty}\frac{z_+^{N_+}z_-^{N_-}}{N_+!N_-!}
\int\{dr^{(+)}\}_{N_+}\{dr^{(-)}\}_{N_+}e^{\beta NV(0)/2} \int\frac{D\chi}{%
{\cal N}}e^{-\beta/2(\chi,H,\chi)} e^{i\beta\int\rho_q(r)\chi(r)dr}
\end{equation}
We will suppose our system to be "neutral", so that $z_+=z_-=z$.
Keeping in mind Eq.(\ref{rho})  we get:
\begin{equation}
\label{Hgordon}
\zeta=\int\frac{D\chi}{{\cal N}}\exp\left(-\beta/2(\chi,H,%
\chi)+ 2\tilde z\int\cos\beta\chi(r)dr\right),
\end{equation}
where $\tilde z=ze^{\beta V(0)/2}$. For the case of Coulomb gas, where
$V(q)\sim \frac{1}{q^2}$, and $H(r-r')\sim\delta(r-r')\nabla^2$ the equation
(\ref{Hgordon}) becomes similar to the Euclidean Sine-Gordon theory.
This equivalence is a well known fact for more than two
decades~\cite{samuel,minnh87}.
To calculate approximately $\zeta$ we can use a perturbation theory in the
spirit of quantum field theory \cite{ramon}. For this purpose we need to
consider again:
\begin{equation}
Z[J]=\int\frac{D\chi}{{\cal N}}e^{-\beta/2(\chi,H,\chi)}e^{\int J\chi dr}
=\exp\left(\frac{1}{2\beta}\int
J(r)V(r-r^{\prime})J(r^{\prime})drdr^{\prime}\right),
\end{equation}
and to define:
\begin{equation}
\label{mean}
<f(\chi(r))>_H=
\int\frac{D\chi}{{\cal N}}f(\chi(r))e^{-\beta/2(\chi,H,\chi)}.
 \left.f(\frac{\delta}{\delta J(r)})Z[J]\right|_{J=0}.
\end{equation}
In this case we obtain for $\zeta$:
\begin{equation}
\label{funcderiv} \zeta=\left.\left<\exp\left(2\tilde
z\int\cos\beta\chi(r)dr\right)\right>_H= \sum_{n=1}^{\infty}\frac{1}{n!}%
\left[ 2\tilde z\int\sum_{k=0}^{\infty}\frac{(-1)^k}{(2k)!} \left( \beta
\frac{\delta}{\delta J(r)} \right)^{2k}dr\right]^nZ[J]\right|_{J=0}.
\end{equation}

Following methods widely used in quantum field theory, we represent each
 term of the sum in (\ref{funcderiv}) as a diagram with $n$ vertex connected by
$k$ lines. It is easy to show that an even number of lines are closed
in each vertex. Each vertex ($i$) contributes an integration over $dr_i$,
the factor $-\beta V(r_{ij})$ corresponds to each line connecting vertices
$i$ and $j$.

Quantum field theorists know that $\zeta$ (Eq. (\ref{funcderiv})) is infinite
in the limit $\Omega\to\infty$. The infiniteness of $\zeta$
(Eq. (\ref{GrandPartSum})) is also known from statistical physics.
To our good both cases need only the value of $\ln\zeta/\Omega$.
This permits us to exclude all disconnected diagrams, like one usually
acts in quantum field theory.

        Now it's time to examine diagrams, containing tadpoles.
Tadpole is a line $(-\beta V(r_{ii}))$ which begins and ends at the same
vertex, so that the resulting muliplying factor is proportional to $V(0)$.
The term "tadpole" is well known by field theorists, but is used very seldom
in classical statistical mechanics.

        Let us consider a vertex  with $l$ tadpoles and $j$ non-tadpoles
lines outgoing from it (external lines with respect to vertex). The
combinatorial prefactor in the integral, corresponding to the diagram with such a tadpole is:
$$
\frac{(2k)!}{l!j!2^l},
$$
where $2k=2l+j$ is a full number of lines outgoing from the vertex.
(note that $j$ must be even)

Summing by  a number of tadpoles $l$ of the vertex with fixed number $j$ of
external lines,
we can prove, that all tadpoles of the vertex can be absorbed into
a factor  $e^{-\beta V(0)}$ bound to this vertex.
This is equivalent to the renormalization   of
$\tilde z$  to its original value:
\begin{equation}
\label{re-re}
\tilde z=e^{\beta V(0)/2}z\mapsto e^{-\beta V(0)/2}\tilde z=z.
\end{equation}

Now the role of the factor $e^{N\beta V(0)}$ in Eq. (\ref{gibbsgauss})
becomes clear. The existence of this factor enables us to compensate
the singular contributions of tadpoles in diagrams. To compensate
these tadpoles in quantum field theory the normal ordering of the initial
Gamiltonian is usually used (see for example~\cite{BSH}). In the case
of statistical mechanics we have
no base neither to use normal ordering, nor to suppose our
initial parameters to be infinite, so that the singular factor appearing in
Eq. (\ref{gibbsgauss}) is the only mathematically correct way to avoid
singularities bound with tadpoles.

At the same time it is important to note that such a re-renormalization
(\ref{re-re}) takes place only if we consider the full interaction
$\cos\beta\varphi$.\footnote{Here the quantum field theory terminology
is used: $(\chi,H,\chi)$ -- is a Hamiltonian's kinetic term, while
($cos(\beta\varphi)$) is the interaction.} In the case of substitution:
$$
\cos\beta\varphi=1-\frac{(\beta\varphi)^2}{2}+\frac{(\beta\varphi)^4}{4!},
$$
the maximum number of lines in the vertex becomes equal to 4, and the
maximum number of tadpoles is  limited, so that the factor $e^{\beta V(0)}$
can not be fully compensated by them.

After the discussion of tadpole-compensating it is not
difficult to show by analogy, that summing of multiple lines, connecting
two vertexes allows us to change each set of diagrams with multiple
lines by one diagram  with "bold" line, where
bold lines are assosiated with Mayer's function $f_{ij}=e^{-\beta V(r_{ij})}-1$.
Here we shall not concentrate on the proof of this fact. The analogous
result is described in \cite{balesku}.

        The main conclusion from the above consideration is that the
perturbation theory derived from the path integral representation
of the Grand partition sum (see Eq. (\ref{Hgordon})) -- path integral
perturbation theory, is equivalent to the Mayer's
series derived directly from Eq. (\ref{GrandPartSum}), and
can be written through the same parameters: $z, f_{ij}$.

        At the same time, despite the similarity of these two evaluation,
there are a difference between them: really, Mayer's diagrammatic expansion
of (\ref{GrandPartSum}) contains two kind of vertices, corresponding
to both (positive and negative) kinds of particles. Factors $z_+$ and
$z_-$
correspond to these vertices. Each pair of vertices can be connected by two
kind of lines, depending  on their mutiple charges:
$e^{-\beta V_{+-}(r)}$ and $e^{-\beta V_{\pm\pm}(r)}$. To understand
why  $f_{+-}$-type lines do not appear in the path integral
perturbation theory, we can expand Mayer's function
$f_{ij}=e^{-\beta V_{ij}}-1$,
in powers of $V_{ij}$. The difference between  $f_{++}$ and $f_{+-}$
will be seen only in diagrams, containing one or more vertices with even
number of lines (even vertices). For each diagram containing even vertex
we can find another graph, differ only in a "sign" of this even vertex.
 Due to the neutrality of our system ($z_+=z_-$ -- see above) these
diagrams will compensate with each other, giving no contibution to the
final free energy. Thus we have shown that for Mayer's developement of the
grand partition sum, after representation of $f_{ij}$ as  series (see for
example \cite{balesku}) diagrams with  even vertices do not contribute
to a final result. Summing the rest of diagrams we get the resut identical
to the one obtained from the path integral perturbation theory.

In this section we have shown schematically the similarity and moreover,
the equivalence of both perturbation theories. This result gives us
no new methods to calculate thermodynamic
properties of the system, but is important for understanding
different approximations obtained from Eq.(\ref{Hgordon}) or directly
from Eq. (\ref{GrandPartSum}). This is  not a purpose of this section
to perform a full consideration of different perturbation theories
connected with Mayer's series.
 Here we consider only those results that can by obtained
from the path integral representation of the partition sum.

\section{The expansion in  powers of density}
\label{density}

To obtain the expansion in powers of density, it is  convenient to consider
the partition sum instead of the grand partition sum, used in the previous
section.
For the system of two sorts of particles with densities $\rho_+$ and
$\rho_-$ the partition sum can be written as:
\begin{equation}
\label{partsum}
Z=\frac{{\lambda_+}^{N_+}}{N_+!}\frac{{\lambda_-}^{N_-}}{N_-!}
\int\{ dr\}\exp\left\{-\frac{\beta}{2}\sum_{ij}V_{ij}(r_{ij})\right\}.
\end{equation}
Here $\{dr\}$ denote the integration over all $N_++N_-$ particle positions
in $d$-dimensions, $\lambda=\left(\frac{K_bTm}{2\pi\hbar}\right)^{1/2}$ is
a Boltzmann wavelength. We obtain the free energy $F$:
$$
-\beta F=\ln Z= N_+\left[\ln\left(\frac{\lambda_+}{\rho_+}\right)+1\right]+
N_-\left[\ln\left(\frac{\lambda_-}{\rho_-}\right)+1\right]+\ln Q,
$$
where $\ln Q$ is the nonideal part of  the free energy:
\begin{equation}
\label{lnQ}
\ln Q(N_+,N_-,\beta,\Omega)= \ln\int\frac{dr_1}{\Omega}\cdots
\frac{dr_N}{\Omega}e^{-\beta U},
\end{equation}
$\Omega$ is the $d$-dimensional volume of the system.

Now we rewrite $\exp(-\beta U)$ using Eq. (\ref{gibbsgauss}):
\begin{equation}
\label{Qdh}
Q=\exp(\frac{N\beta}{2}V(0)) \int\frac{D\chi}{{\cal N}}
\exp\{-\beta/2(\chi,H\chi)\}e^{S_1},
\end{equation}
$$
S_1=N_+\ln\int\frac{dr_+}{\Omega}e^{i\chi(r_+)}+ N_-\ln\int\frac{dr_-}{%
\Omega}e^{-i\chi(r_-)}
$$
To calculate the path interal $D\chi$ we need to diagonalize the
quadratic  form in the exponent of Eqs. (\ref{Qdh}),(\ref{S_1}). For this
purpose it is convenient to use Fourier transform as a change of variables
in path interals \cite{feinman}
\begin{equation}
\chi(r)=\int\frac{dq}{(2\pi)^d}\hat\chi_qe^{-i{\bf qr}},\qquad
\hat\chi_q=\int dr\chi(r)e^{i{\bf qr}}.
\end{equation}
$\chi(r)$ is real function, its Fourier transform satisfies the
requirements
$\hat\chi_q^*=\hat\chi_{-q}$.

For our further consideration we suppose $\chi$ to oscillate around
$\chi\equiv 0$, so that $S_1$ can be expanded just to the second
order in $\chi$.
In the case of neutrality ($\rho_+=\rho_-$) we get:
\begin{equation}
\label{S_1}
S_1=\frac{\rho_++\rho_-}{\Omega}|\hat\chi_0|^2- \frac{\rho_++\rho_-}{2}\int
\frac{dq}{(2\pi)^d}|\hat\chi_q|^2.
\end{equation}

Path integration over $D\chi(r)$ should be replaced by:
$$
D\hat\chi_q\equiv d\hat\chi_0\prod_{q>0}d{\rm Re}\hat\chi_qd{\rm Im}
\hat\chi_q,
$$
where $q>0$ denotes the production through a "half"-space of $q$. Here it
is important to remind the mathematical definition of the path integral as
a limit of multiple integral over function values on the discrets set of
points.
Keeping in mind the simmetry of $V(r)$,and $H(r)$ we can write:
\begin{equation}
\int\chi(r)drH(r-r^{\prime})dr^{\prime}\chi(r^{\prime})=\int\frac{dq}{%
(2\pi)^d}|\chi_q|^2\hat H(q).
\end{equation}
In this case for the exponent in Eq.(\ref{Qdh}) we get:
\begin{equation}
\label{s1}
-\beta/2\sum_{q>0}\frac{2}{\Omega}(\xi_q^2+\tilde\xi_q^2)\hat
H(q)- \beta^2\frac{\rho_++\rho_-}{2}\sum_{q>0}\frac{2}{\Omega}(\xi_q^2+
\tilde\xi_q^2)-\beta/2\frac{1}{\Omega}\hat H(0)\hat\chi_0^2.
\end{equation}
Here we use the summation instead of the integration over $q$-space following the
rule:
$$
\int\frac{d{\bf q}}{(2\pi)^d}=\frac{1}{\Omega}\sum_q.
$$
The notation $\sum\limits_{q>0}$ means a summation over "half"-space
$\cal D$ of the $q$-space , so that for each $q$ in $\cal D$, $-q$ is not
in $\cal D$. Considering this sum we specially marked a point $q=0$.
We also introduced for convenience:
\begin{equation}
\label{xidef}
\xi_q={\rm Re}\chi_q\equiv{\rm Re}\chi_{-q};\qquad\tilde\xi_q=
{\rm Im}\chi_q\equiv-{\rm Im}\chi_{-q};\qquad \chi_0=\chi_0^*.
\end{equation}

Now we can represent the configuration integral $Q$ using the above formulas:
\begin{eqnarray}
\label{Qdh1}
Q&=&\exp\left(\frac{N\beta}{2}V(0)\right)\frac{1}{\cal N}\int\prod_{q>0}
d\xi_qd\tilde\xi_qd\chi_0\times\nonumber \\
&\times&\exp\left[\frac{1}{\Omega}\sum_{q>0}\left(-\beta\hat H(q)-
\beta^2(\rho_++\rho_-)
\right)(\xi_q^2+\tilde\xi_q^2)-
\left(\frac{\beta}{2\Omega}\hat H(0)\chi_0^2
\right)\right].
\end{eqnarray}
Analogously, the  normalizing factor $\cal N$ can be written as:
\begin{eqnarray}
{\cal N}&=&\int\prod_{q>0}d\xi_qd\tilde\xi_qd\chi_0\times\nonumber \\
&\times&
\exp\left[
\frac{1}{\Omega}\sum_{q>0}
        \left(\beta\hat H(q)(\xi_q^2+\tilde\xi_q^2)\right)
        -\frac{\beta}{2\Omega}\hat H(0)\right]
\end{eqnarray}

After simple calculations, we obtain for $\ln Q$:
\begin{equation}
\ln Q=\frac{N\beta}{2}V(0)+ \sum_{q>0}\ln\frac{\beta\hat H(q)}{\beta\hat
H(q)+\beta^2(\rho_++\rho_-)}.
\end{equation}
At the final stage the sums should be replaced back by the integrals:
\begin{equation}
\label{DHpure}
\ln Q=-\frac{\Omega}{2}\int\frac{dq}{(2\pi)^d}
\ln(1+\beta(\rho_++\rho_-)\hat V(q))+\frac{(\rho_++\rho_-)\beta\Omega}{2}
\int\frac{dq}{(2\pi)^d}\hat V(q).
\end{equation}
The last term in Eq. (\ref{DHpure}) reflects the fact that:
$$
V(0)=\int\frac{dq}{(2\pi)^d}\hat V(q).
$$
Finally, for the nonideal part of the free energy we obtain:
\begin{equation}
\label{respure} \frac{\beta\Delta F}{\Omega}=\frac{1}{2}\int\frac{dq}{%
(2\pi)^d}\left\{ \ln(1+\beta(\rho_++\rho_-)\hat
V(q))-\beta(\rho_++\rho_-)\hat V(q) \right\}.
\end{equation}

For the case of the neutral 3D plasma  (\ref{coulomb}), this equation
is known as a Debye-Hukkel approximation \cite{balesku} for the free
energy. The way, presented above
ia a new method to obtain the well known result directly from the Gibbs
partition sum without special physical suppositions.

 The fact that Eq.(\ref{respure}) is identical to the result of the summation of cycles
diagrams can be simply explained. Really, the expansion of $S_1$ in
(\ref{Qdh}) just to the second order of $\chi$ corresponds
to the retaining for later summation only  those diagrams which have no
more than
two lines in each vertex. The vertices without lines appears in disconnected
diagrams only, graphs with one line  vertexes compensate themselves if
the system
is neutral. The only diagrams leaving for the summation are cycles, and
their summation leads to Eq. (\ref{respure}).

       At the same time, the path integral method described above
explains us, why the result (\ref{respure}) cannot be applied to
the non-Coulomb like systems. For the Van-der-Vaals interparticle interaction
$V(r)$, the Fourier transform $V_q$ is usually not positively defined (see
section \ref{gau}). Due to this fact  we cannot apply all the above
considerations to the non-Coulomb-like systems. To obtain the analogue of
Eq. (\ref{respure}) in a general case, we need to invent the analogue of
the Gauss representation (\ref{discr}) with a sign-alternating quadratic
form $A_{ij}$ and to extend it to the continuous case.
 The aim  of this paper is not to realize this plan, but to show
some features of  the path integral method in statistical
mechanics, and to obtain some new results for the simple case of Coulomb
gas.

\section{System with dipolar pairs}
\label{dipols}

The interesting chance appears while using the path integral representation of
the Gibbs partition sum: we can include clusters of particles in our consideration.
For simplicity let us consider a system, consisting of $N_+^F$ positively
charged and $N_-^F$ negatively charged particles.  A part of these particles
is organized in clusters of known form. The intercluster interaction
and their self energy can be calculated by summing "each with each"
interparticles interactions. This fact can be easily used to
include clusters of particles in Eq.(\ref{gibbsgauss}).

As a simple example we consider a system of $N_++N_-$  positive and negative
free particles. We will  suppose also that there are $N_D$ dipoles.
Here a dipole is considered as a pair of particles ($+$, $-$) with distance
between them
to be fixed and equal to $a$. Such a dipol has $2d-1$ degrees of freedom.
We can write the microscopic charge density $\rho_q$ of the system as:
\begin{equation}
\rho_q=\sum_{i=1}^{N_+}\delta(r-r^+_i)-\sum_{j=1}^{N_-}\delta(r-r^-_j)+
\sum_{k=1}^{N_D}(\delta(r-r^{D_+}_k)-\delta(r-r^{D_-}_k)),
\end{equation}
where $r_+$ and $r_-$ are positions of the free particles and
$r^{D_{\pm}}=r^D\pm{\bf a}/2$ are  positions  of positive and negative
charges of the dipole.

Gibbs partition sum for the system is:
\begin{equation}
\label{Z_3}
Z_3=\frac{{z^{N_+}}_+}{N_+!} \frac{{z^{N_-}}_-}{N_-!}\frac{{z^{N_D}}_D}{N_D!}
\Omega^{N_++N_-+N_D}\int \left\{\frac{dr^+}{\Omega}\right\}_{N_+} \left\{
\frac{dr^-}{\Omega}\right\}_{N_-} \left\{\frac{dr^D}{\Omega}\frac{d\omega}{%
\omega_0} \right\}_{N_D}\exp\left\{-\beta U\right\}.
\end{equation}
Here $z_{\pm}$ are activities of free particles and $z_D$ -- activity  of
dipoles\footnote{Here we consider dipoles as the third kind of particles}.
We integrate in (\ref{Z_3}) over positions
of all free particles ($dr^\pm$), and over positions of dipol's centers
($dr^D$). In addition in Eq. (\ref{Z_3}) the integration over all orientations
of dipols $\left(\frac{d\omega}{\omega_0}\right)$ where $\omega_0$ is a
normalizing factor is performed.
For the case $d=2$ we have
$\frac{d\omega}{\omega_0}=\frac{d\varphi}{2\pi}$, for $d=3$ --
$\frac{d\omega}{\omega_0}=\frac{\sin\theta d\theta d\varphi}{4\pi}$.

Taking into account the "each-with-each" interaction of
particles, we write for $U$:
\begin{equation}
U=\sum_{i\ne j}^{N_++N_-+2N_D}s_is_jV(r_{ij}),
\end{equation}
so that $U$ includes not only ions-dipoles and dipoles-dipoles interactions,
but interaction of particles inside the dipoles. This fact is important when
we define the activity of the dipoles.

Due to Eq.(\ref{gibbsgauss}) we represent $Z_3$ through the path integral:
\begin{equation}
Z_3=\exp\left\{\frac{N\beta V(0)}{2}\right\} \int\frac{D\chi}{{\cal N}}%
\exp(-\beta/2(\chi,H\chi)+S_2) \frac{{z^{N_-}}_-}{N_-!}\frac{{z^{N_D}}_D}{%
N_D!}\Omega^N.
\end{equation}
Where $N=N_++N_-+2N_D$, and
\begin{equation}
S_2=N_+\ln\int\frac{dr}{\Omega}e^{i\beta\chi(r)}+ N_-\ln\int\frac{dr}{\Omega}%
e^{-i\beta\chi(r)}+ N_D\ln\int\frac{dr}{\Omega}\frac{d\omega}{\omega_0}
e^{i\beta\left(\chi(r_D+a/2)-\chi(r_D-a/2)\right)}
\end{equation}

Now we expand $S_2$ just to the second order of $\chi$, as done
in section \ref{density}. For convinience we use the Fourier transform
of $\chi(r)$. Using:
$$
\chi(r_D+a/2)-\chi(r_D-a/2)=-2i\int\frac{dq}{(2\pi)^d}\hat\chi(q)e^{-iqr_D}
\sin({\bf qa}/2),
$$
we  get for $S_2$:
\begin{equation}
S_2\approx\frac{\rho_++\rho_-}{2}\left[ \frac{\hat\chi_0^2\beta^2}{\Omega}-
\beta^2\int\frac{dq}{(2\pi)^d}|\hat\chi_q|^2 \right]-2\rho_D\beta^2\int\frac{%
dq}{(2\pi)^d}|\hat\chi_q|^2\Gamma(q),
\end{equation}
where a function:
\begin{equation}
\label{Gamma} \Gamma(q)=\int\frac{d\omega}{\omega_0}\sin^2(\frac{{\bf qa}}{2}%
)
\end{equation}
is used.
In Eq.(\ref{Gamma}) the integration is over all directions $\omega$ of
vector ${\bf a}$.

Using the same notations as in section \ref{density} we can perform
the same sequence of calculations and for the exponent (similar to
Eq. (\ref{s1}))
we can write:
$$
\label{s2}
-\beta/2\sum_{q>0}\frac{2}{\Omega}(\xi_q^2+\tilde\xi_q^2)\hat
H(q)-
\beta^2\frac{\rho_++\rho_-}{2}\sum_{q>0}\frac{2}{\Omega}(\xi_q^2+
\tilde\xi_q^2)-
4\beta^2\rho_D\frac{1}{\Omega}\sum_{q>0}(\xi_q^2+\tilde\xi_q^2)\Gamma(q)-
$$
\begin{equation}
\qquad \beta/2\frac{1}{\Omega}\hat H(0)\hat\chi_0^2-2\beta^2\rho_D
\frac{1}{\Omega}\Gamma(0)\hat\chi_0^2.
\end{equation}
The configuration integral can be written as:
\begin{equation}
Q=\exp\left(\frac{N\beta}{2}V(0)\right)\frac{1}{{\cal N}}\int\prod_{q>0}
d\xi_qd\tilde\xi_qd\chi_0 \exp\left\{\frac{1}{\Omega}\sum_{q>0}\left[-\beta
H_q-\beta^2(\rho_++\rho_-)- \right.\right.
\end{equation}
$$\left.\left.
4\beta^2\rho_D\Gamma(q)\right]\left(\xi_q^2+\tilde\xi_q^2\right)+
\chi_0^2\left(-\frac{\beta}{2\Omega}H(0)-\frac{2\beta^2\rho_D}{\Omega}
\Gamma(0)\right)\right\}.
$$
Here  ${\cal N}$ -- is the same normalizing factor as in section \ref{gau}.
Finally we obtain for $\ln Q$:
\begin{equation}
\label{LnQDipol}
-\beta F_d=\ln Q=-\frac{1}{2}\ln(1+4\beta\rho_D\Gamma(0)V(0))-
\end{equation}
$$
\Omega\frac{1}{2}\int\frac{dq}{(2\pi)^d} \left\{\ln(1+\beta(\rho_++%
\rho_-)V_q+ 4\beta\rho_D\Gamma(q))-(\rho_++\rho_-+2\rho_D)\Omega
V_q/2\right\},
$$
where the first term  can be neglected as it remains finite in the limit
$\Omega\to\infty$

Now we can review additional suppositions introducted in this section while
obtaining Eq. (\ref{LnQDipol}). The only special supposition in comparison with
section \ref{density} is the definition of dipoles and their including in
the partition sum (\ref{Z_3}). Other derivations, mathematically are
fully equivalent to ones made in section \ref{density}. Moreover, it is
obvious that in the limit $\rho_D\to 0$
Eq.(\ref{LnQDipol}) reduces to the nonideal part of the free energy of
pure two-components plasma Eq.(\ref{respure}).  This fact makes us to hope
that the result (\ref{LnQDipol}) can be considered as a generalization of the
Debye-Hukkel result for a two compomemt plasma with dipoles.

The including of dipole pairs is not the only way to get some new analytical
results
using the path integral formulation in statistical mechanics.
  For example to describe a hexatic phase it may be useful to introduce
hexagonal clusters, so that the charge density becomes:
\begin{equation}
\rho(r)=\sum_{i=1}^{N_+}\delta(r-r^+_i)-\sum_{j=1}^{N_-}\delta(r-r^-_j)+
\sum_{k=1}^{N_H}\rho_{ah}(r-r_k),
\end{equation}
Where  $\rho_{ah}$ -- is a local charge density
of the  hexagonal cluster of size $a$:
$$
\rho_{ah}(r-r_k)=s_0\delta(r-r_k)+\sum_{n=1}^6s_n\delta(r-r_k+a_n).
$$
($s_i$) are charges of particles of clusters considered.

In this case we shall get a function $\Gamma(q)$ that is different from
Eq.(\ref{Gamma}). This function will be closely related to the symmetric
properties of clusters.

Analogousely we can introduce a partition sum including many different
types of
clusters. The resulting free energy can be considered as a function of
their densities. We may hope to obtain the physical properties
of our system by minimizing this free energy.
 For example, the free energy  $F_d$  (Eq. (\ref{LnQDipol}))
 can be considered as a function of free particles
($\rho_+\sim\rho_-$) and dipoles ($\rho_D$) densities.
If the minimum of such an energy is placed in the region ($\rho_D\gg\rho_+$),
we can conclude that under these conditions  free particles couple
into dipole pairs.

\section {Conclusion}
\label{conclusion}

In the present paper a new point of view on the classical Gibbs partition
sum is presented. This method is based on the representation of the
configuration integral through the Gaussian path integration. Obtained
results precisely reproduce the Debye-Hukkel result  for the symmetric
3D Coulomb plasma.
As it is shown, this method and its extensions can be applied to the case
of Coulomb and quasi-Coulomb systems.
Presented method is applyed  to the very interesting case of symmetric
Coulomb plasma with dipoles~\cite{fisher} and the genralization of DH
result was obtained.

Presented point of view on the statistical mechanics has some advantages
and defects.
The main defect of this point of view is a formality of computations.
Really, all transformations presented above are simple extensions
of the well known techniques of multiple Gaussian integrals computation.
By the way  we were forced to ommit the analysis of some questions
concerning the divergence of intermediate results.

Talking about the advantages of the described method, it should be mentioned
that we got a  method  allowing us to find the range of the applicability
 of the DH result (see section \ref{gau}). Also we can see
possibles systematic ways to obtain higher order  correction to the known results.
For example it is obvious that the next order of the DH result is the
well known corrections to the Gaussian approximations of the integral:
techniques of calculation of these corrections is well developped,
and we can try to generalize it to the continuum case.
Analogousely, we can  extend  DH (or DH-like) result for the case of
more complex systems with Coulomb interparticle interaction,
as is done in sec. \ref{dipols}.

Method presented in this paper should be extended and analyzed
from differents points of view:
First of all it should be interesting to obtain numerical results using
new formulas derived above and to compare these calculations with
others approximations.
We hope to get new results by applying the path integral representation
to Coulomb systems with differents types of clusters,as it is done with
dipols in section \ref{dipols}.
As it was pointed above, the generalization of the Gaussian path integration
to the case of non positively defined Fourier  transforms is needed to
widely test the presented  method.
 Another topic of interest is the connection of the
path integral representation  with the renormgroup approach and critical
phenomena. There is a good chance that the presented non-trivial
representation of the Gibbs partition sum can give us new approaches and
methods to physics of critical phenomena.

\section{acknowledgments}

I would like to thank E. E. Tareyeva and V. N. Ryzhov for interesting 
discussions and for reading the manuscript.
This work is supported in part by the Russian Science Foundation through 
the Grant N96-02-16211.


\end{document}